\documentclass[10pt,aps,prd, floats, floatfix, superscriptaddress, twocolumn,nofootinbib,amsmath,amssymb]{revtex4-2}

 \usepackage{enumerate}
 \usepackage{graphicx}
 \usepackage{multirow}
 \usepackage{xcolor}
 \usepackage{tablefootnote}
\usepackage{threeparttable}
 \usepackage{bm}
 \usepackage[normalem]{ulem}

 \usepackage{comment}

\usepackage{amsmath}
\usepackage{enumerate}
\usepackage{amssymb}
\usepackage{graphicx}
\usepackage{multirow}
\usepackage{xcolor}
\usepackage{tablefootnote}
\usepackage{threeparttable}
\usepackage{bm}
\usepackage[normalem]{ulem}
\usepackage{physics}
\usepackage{cancel}
\usepackage{soul}
\usepackage{tikz}
\usepackage{braket}
\usepackage{simpler-wick}
\usepackage{amsmath}
\usetikzlibrary{decorations.pathreplacing}

\usepackage{amsmath,mathtools}
\usepackage{float}

\usepackage[acronym]{glossaries}
 \usepackage{physics}

 \newacronym{frw}{FRW}{Friedmann-Robertson-Walker}
 \newacronym{cmb}{CMB}{Cosmic Microwave Background}

\usepackage{soul}

\begin{document}

\title{A Bell experiment during inflation: probing quantum entanglement in tensor fluctuations through correlations of primordial scalar curvature perturbations}

\author{Pablo Tejerina-P\'erez}
\affiliation{ICC, University of Barcelona, Mart\' i i Franqu\` es, 1, E08028
Barcelona, Spain}

\author{Leonid Sarieddine}
\affiliation{ICC, University of Barcelona, Mart\' i i Franqu\` es, 1, E08028
Barcelona, Spain}

\author{Daniele Bertacca}
\affiliation{Dipartimento di Fisica e Astronomia Galileo Galilei, \\Universit\`a degli Studi di Padova, via Marzolo 8, I-35131, Padova, Italy}
\affiliation{INFN, Sezione di Padova, via Marzolo 8, I-35131, Padova, Italy}
\affiliation{INAF- Osservatorio Astronomico di Padova, \\ Vicolo dell Osservatorio 5, I-35122 Padova, Italy}

\author{Raul Jimenez}
\affiliation{ICC, University of Barcelona, Mart\' i i Franqu\` es, 1, E08028
Barcelona, Spain}
\affiliation{ICREA, Pg. Lluis Companys 23, Barcelona, 08010, Spain.}

\begin{abstract}
We propose a method that provides an observational signature of the quantum origin of primordial fluctuations generated during inflation. The method gives a prescription for testing a Bell inequality constructed exclusively from the standard scalar and tensor perturbations of minimal single-field inflation. We consider an inflationary spacetime populated by pairs of gravitons entangled in their polarization states. Third-order interactions between two scalars and one graviton transfer polarization information to the scalar sector through the product of spatial derivatives of scalars with the tensor polarization factors. Rather than performing the full multidimensional momentum integrations, we isolate and compute the tensor polarization structure of the primordial scalar eight-point correlation function. This eight-point correlation function factorizes into the product of four scalar two-point functions associated with opposite (mirrored) momentum configurations in Fourier space. This factorization falls from the fact that the two gravitons are spatially well-separated within the cosmological horizon of inflation, replicating the setup of standard Bell experiments. Through these interactions, we track how non-local correlations between both gravitons from polarization entanglement are imprinted on the scalar sector. We show that, for specific configurations of the scalar momenta after the end of inflation (detailed in the text), this observable can be used to construct a Bell-violating quantity in a way that matches the well-known Clauser-Horne-Shimony-Holt inequality definition.  In principle, this offers a route to probe the quantum nature of primordial fluctuations through observables accessible today.

\end{abstract}

\maketitle

\section{Introduction}

Inflation is a paradigm that posits a period of exponentially accelerated expansion of spacetime, which can be well approximated by a four-dimensional de Sitter spacetime with a slightly broken time-translational symmetry that allows for a graceful exit into the Friedmann–Lemaître–Robertson–Walker metric (FLRW) we observe today (see e.g. ~\cite{Mukhanov_1} ) 

It was first introduced by ~\cite{Guth1981,Sato1981,Linde1981,Starobisnky_inflation1982,Albert_Steinhard1982}. It is today the most widely accepted theory of the beginning of the Universe, since it solves naturally the shortcomings of the ``Big Bang'' model, such as the horizon or flatness problems (see e.g. \cite{Mukhanov_1, Dodelson2003ft}). Further, inflation naturally introduces a mechanism for the production of primordial matter density fluctuations that act as seeds for the formation of the large-scale-structure (LSS) and the anisotropies that we observe today~\cite{Mukhanov:1981xt}. These seeds come from quantum fluctuations in the spacetime metric and the fields within it, that get enlarged to cosmological scales by the fast expansion of the universe.

The simplest model of inflation is  \textit{slow-roll, single-field} inflation. It is characterized by the presence of a single scalar field denoted as the \textit{inflaton field}, and the spacetime metric field \cite{Mukhanov_1}. This scalar field couples to the metric through a slightly tilted potential, which gives a non-zero vacuum energy for the background inflaton field that drives the expansion for an appropriate amount of e$-$foldings (number of times the Universe exponentially grows in size). At the end of the inflationary phase, the inflaton vacuum energy gets injected into the rest of species through oscillatory dynamics about the minimum of the scalar potential in an epoch called \textit{reheating} \cite{bassett2006inflation, allahverdi2010reheating}.

In this context, the mechanism for production of fluctuations around the background value of the de Sitter metric and of the inflaton field is just a consequence of the quantum nature of fields. Heisenberg's uncertainty principle implies the appearance of random fluctuations in both fields, that then evolve as dictated by the coupled Einstein-Boltzmann equations \cite{mukhanov1992theory, ma1995cosmological}. During inflation, the fast expansion stretches the wavelength of these fluctuations to cosmological scales. At some time, each wavelength becomes of the order of the cosmological horizon present in de Sitter metric. At this point (and namely in the standard single-field adiabatic case) we say the fluctuation freezes, and its amplitude remains approximately constant while the wavelength keeps on growing. These perturbations of cosmological scale re-enter the cosmological horizon in the post-inflationary phase and present themselves as inhomogeneities in the spacetime metric in an otherwise homogeneous Universe. Matter then undergoes gravitational collapse induced by the perturbations, and the non-linear evolution of Einsteins equations drives the process of structure formation. In this way, the LSS is imprinted with the physics behind the perturbations of inflation. At the largest scales, where the linear regime of Einstein's equations is still valid, this imprinting of the inflationary physics might be accessible \cite{kodama1984cosmological, biagetti2019hunt}.

The treatment of inhomogeneities and anisotropies present in the LSS and the Cosmic Microwave Background (CMB), and their correlations coming from the primordial perturbations generated by inflation is extensively performed classically (see e.g. Ch. 8 in \cite{Dodelson2003ft}, or \cite{Planck2018}). This means, even if the initial conditions for the Einstein-Boltzmann equations coming from inflation are quantum-mechanically generated, post-inflationary statistics do not capture the quantum nature of the fluctuations themselves. This is because these initial conditions could potentially be mimicked by a classical distribution, as discussed in \cite{Polarski_Starobinsky_decoherence_1996, Starobisnky_quantum_to_class_1998, Kiefer:2008ku}. It is a natural question to wonder if there exists some imprint or tracer of the quantum origin of primordial perturbations that can be probed today.

Extensive work has been done in trying to search for quantum signatures of the early Universe \cite{Fano_HBT_interferometry,Campo:2003pa,Campo:2003gb,Campo:2005sv,Campo:2005qn,Campo:2007ar,Campo:2008ju,Campo:2008ij,Choudhury:2016cso,Choudhury:2016pfr,Kanno:2015ewa,Martin:2015qta,Kanno:2017teu,Martin:2017zxs,Martin:2018zbe,Martin:2018lin,Kanno_2019,kanno2020polarized,Danielson:2021egj,Colas:2022kfu,Prabhu:2022zcr,DaddiHammou:2022itk,Ning_2023}. Quantum discord (a measure of ``\textit{quantumness}'') of inflationary perturbations is calculated in \cite{J.Martin_V.Venin} and suggests some features to probe different levels of discordance in CMB descriptions. The effect of entanglement in scalar and tensor fluctuations on CMB power spectra and polarization components $\bf E$ and $\bf B$ is discussed in  \cite{Collins_2016,Bolis:2016vas}. Possible observational signatures in the LSS (clustering of high-redshift, massive collapsed structures) of graviton exchange between tensor and scalar fluctuations are discussed in \cite{Bellomo_2018}. Classicalization of fluctuations during inflation due to squeezing is discussed by Polarski and Starobinsky in \cite{Polarski_Starobinsky_decoherence_1996}. Quantum decoherence, due to the environment or to higher order interactions between tensor and scalar modes, and its traces have also been widely discussed in literature, e.g. see \cite{Calzetta:1995ys, Lombardo:2005iz, Martineau:2006ki, Kiefer:2006je, Kiefer:2008ku, Nelson:2016kjm, Burgess:2014eoa, Martin:2018zbe, Boyanovsky:2015tba, DaddiHammou:2022itk, Burgess_decoherence, Sou:2022nsd, Ning_2023}. 
The most famous experiment regarding quantum entanglement and its non-local nature versus classical correlations is the Bell experiment, introduced in 1964 by John Stewart Bell \cite{Bell}. It has been extensively discussed theoretically and tested in laboratories, see e.g. \cite{Clauser_Shimony, CHSH, Gerry, Walls-Milburn}. In \cite{Maldacena_2015}, J. Maldacena designs what he himself denominates a ``baroque model" of inflation that violates Bell inequalities, which requires an elaborate setup with very specific additional ingredients on top of minimal inflation. 

In our work, we formulate a setup where a Bell inequality can arise within a more realistic model of inflation, where all the necessary ingredients are prescribed in terms of scalar and tensor fluctuations of the metric and a single inflaton field.

This work follows an initial approach presented in \cite{Tejerina2024}, where a mechanism to produce a Bell inequality is formulated by introducing several additional ingredients with respect to minimal slow-roll inflation. The aim of \cite{Tejerina2024} was to explore what extra ingredients additional to minimal single-field inflation could give rise to a maximum signal. In this case, \cite{Tejerina2024} found a signal in the halo bias, which should be measurable in current large scale surveys.

In the present work some key differences are introduced. First, there is a simplification of the measuring mechanism, where we only take the minimal elements given by single-field inflation. 

Second, there is a difference in the choice of observable (although in both cases is a high-order correlation function of scalar perturbations of the metric and the inflaton field), which allows us to give a more explicit construction of the Bell-violating quantity based on this observable. The construction of the latter quantity is inspired and shares common structure with that of \cite{Tejerina2024}. Finally, we give a specific prescription on how our observable could potentially be measured in the LSS of the Universe.

We briefly introduce basic features of our minimal inflationary model in Sec. \ref{sec: Inflation basics}. In Sec. \ref{sec: Outline of a Bell experiment} we summarize the necessary elements for a Bell experiment and schematically describe our prescription for each of this elements in the inflationary setup. Section \ref{sec: realization of Bell experiment} is dedicated to the full formulation: the initial entangled state is presented in \ref{subsec: initial entangled state}, then put into density matrix form for the Bell experiment in \ref{subsec: density matrix formulation}. The relevant physical process is presented and interpreted in section \ref{subsec: our observable - 8 point func}, which gives sense to the choice of observable: an 8-point correlation function of primordial scalar perturbations. This observable is then used in \ref{subsec: CHSH ineq from 8-point} to fully construct a Clauser–Horne–Shimony–Holt (CHSH) inequality that gives a Bell violation. We present our conclusions and future directions in Sec. \ref{sec: Conclusion}.

\section{Inflation basics} \label{sec: Inflation basics}

We take our inflationary model to be slow-roll, single-field inflation. We assume minimal coupling of the massive scalar inflaton field $\phi(x)$, with mass $m$, to gravity (in a spatially flat Friedmann universe). The action for gravity and the scalar field is \cite{Maldacena:2002}:
\begin{equation}
    \label{eq: action 1}
    I = \frac{1}{2}\int d^4x \, \sqrt{g}\,\biggl\{R - g^{\mu\nu}\partial_\mu \phi \, \partial_\nu\phi - 2V(\phi) \biggr\}\,,
\end{equation}

where $V(\phi)$ is the slow-roll potential (i.e. it has non-zero vacuum energy and a small tilt, so the kinetic energy is much smaller than the potential energy, and therefore represents a scalar field with negative pressure \cite{Dodelson2003ft}). We have set $M_p = c = 1$.

The homogeneous solution metric $g_{\mu\nu}$ can be parametrized as:
\begin{equation}
    ds^2 = -dt^2 + e^{2\rho(t)}\, dx_i\,dx^i =  e^{2\rho}\,\left( -d\eta^2 +  dx_i\,dx^i \right)
\end{equation}
where $\eta$ is conformal time $d\eta = e^{-\rho}\,dt$, and we have defined the scale factor $a(t) = e^{\rho}$, so the Hubble parameter is $H=\dot{a}/a = \dot{\rho}$, where the dot represents derivative with respect to cosmic time $t$.

We can express first order perturbations around the background value of the two fields:
\begin{equation}
    \phi(\mathbf{x},t) = \bar{\phi}(t) + \delta\phi(\mathbf{x},t) \,\,\,\,\,;\,\,\,\,\, g_{\mu\nu} = \bar{g}_{\mu\nu} +  h_{\mu\nu}
\end{equation}
where the variables with a ``bar'' represent the background values.\\

Two important gauge choices \cite{Maldacena:2002, Weinberg:2008_cosmology_book, Mukhanov:1988jd, sasaki_1986, baumann2009tasi} are the \textit{comoving gauge}:
\begin{align}
\label{eq: comoving gauge}
    & \delta \phi = 0 \\
    & h_{ij} = e^{2\rho}\, [(1+2\zeta)\,\delta_{ij} + \gamma_{ij}];  \ \ \partial^i\gamma_{ij} = 0;\ \  \gamma^i_{\,\,i}=0 \notag
\end{align}

and the \textit{spatially-flat gauge}:
\begin{align}
\label{eq: spatially flat gauge}
    & \delta \phi = \varphi(\mathbf{x},\eta) \\ 
    & h_{ij} = e^{2\rho}\, (\delta_{ij} + \gamma_{ij})\,; \ \   \partial^i\gamma_{ij} = 0, ; \ \ \gamma^i_{\,\,i}=0 \notag
\end{align}

The physical degrees of freedom are: the gauge invariant variable $\zeta$, which parametrizes the scalar (curvature) fluctuations, and $\gamma$, which parametrizes the tensor perturbations. From now on $\varphi$ represents the fluctuations around the background value of the inflaton field, $\delta\phi = \varphi$.

In section \ref{subsec: our observable - 8 point func}, we will use the spatially flat gauge. The reason will become clear when introducing the relevant action. 

\section{Outline of our Bell experiment during inflation} \label{sec: Outline of a Bell experiment}

\begin{figure}
    \centering
    \includegraphics[trim=45 45 200 80, clip, width=0.99\linewidth]{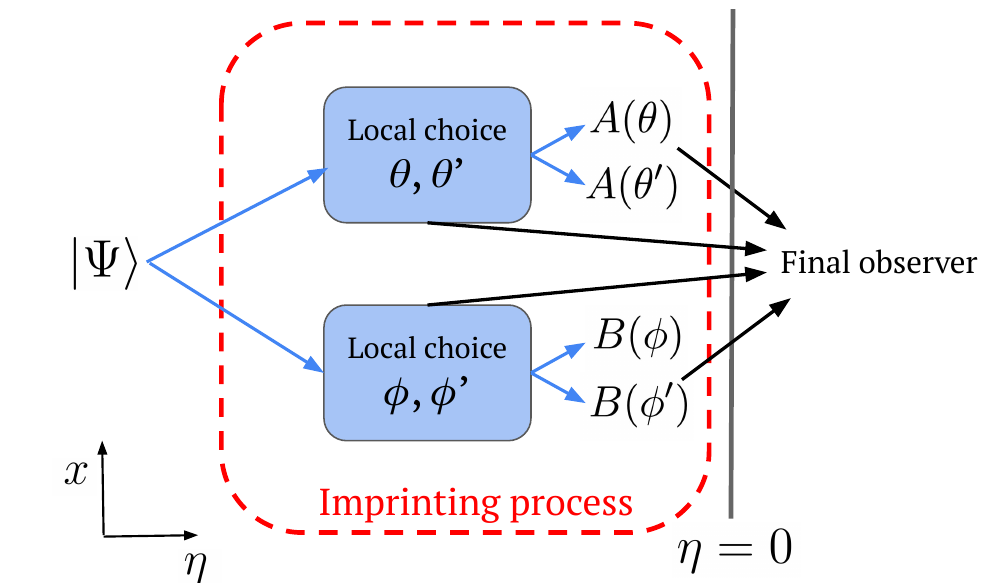}
    \caption{Scheme of the elements and process needed for a Bell experiment during inflation. Conformal time is denoted as $\eta$, being the end of inflation at $\eta=0$. Blue arrows represent the quantum-informed processes. Black arrows correspond to classic evolution and transmission of the results of the quantum experiment to a common location, where they can be correlated by the final observer. The red, dashed box denotes that the imprinting of the quantum correlations needed for a Bell violation should happen after the quantum state has components at spatially separated locations A and B, and before the end of inflation. Figure taken from our previous work \cite{Tejerina2024}.}
    \label{fig: bell experiment scheme}
\end{figure}

\begin{figure}
    \centering
    \includegraphics[trim=160 20 155 20, clip, width=0.8\linewidth]{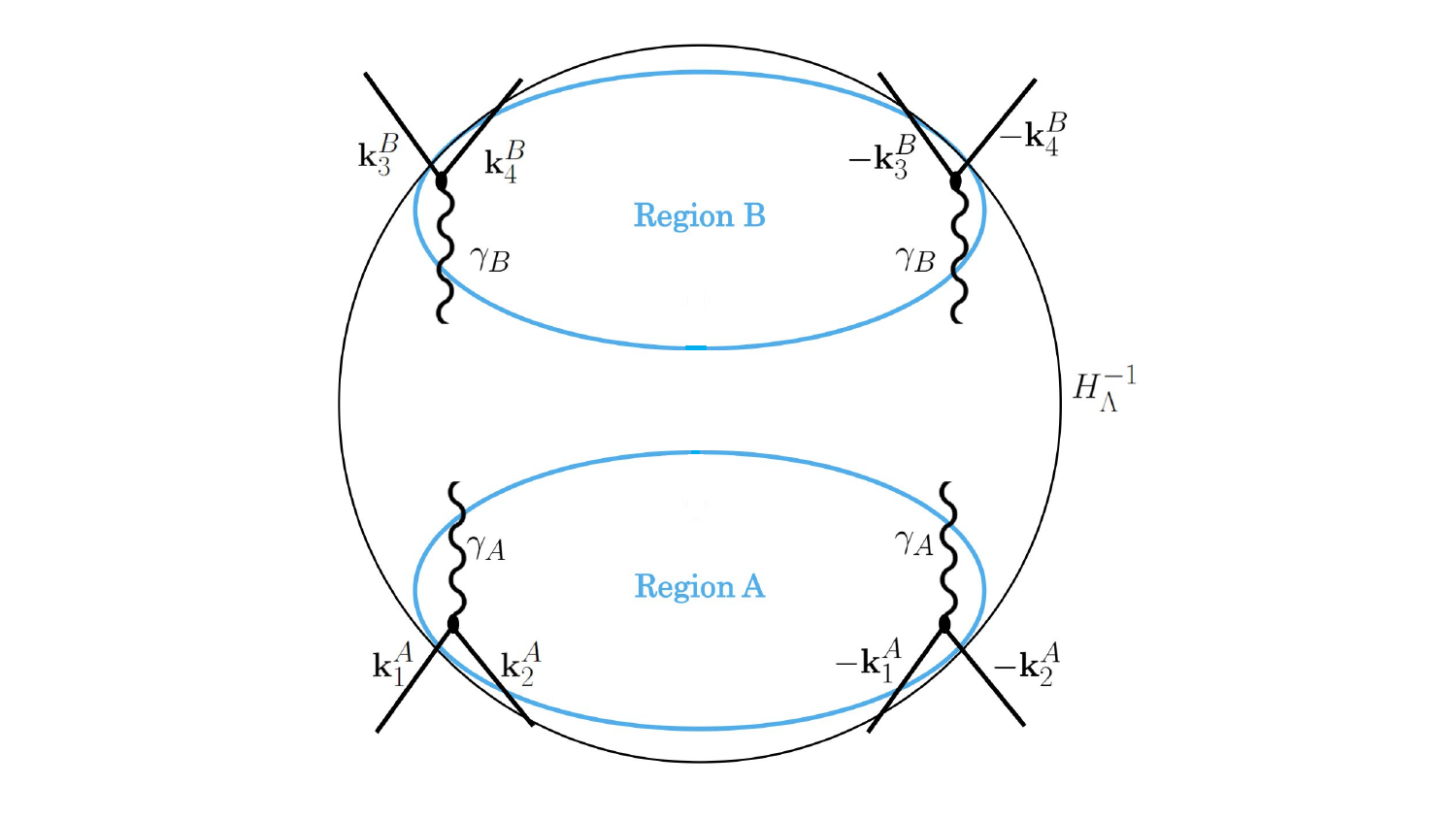}
    \caption{Representation of the relevant process of this work. We depict the interaction vertices corresponding to two gravitons $\gamma_A$ and $\gamma_B$, each interacting twice, with two different pairs of scalar fluctuations. All happens within one Hubble patch of radius $H_\Lambda^{-1}$. Graviton $A$ interacts once with two scalar fluctuations with momenta $\mathbf{k}_1$ and $\mathbf{k}_2$, and once with two scalar fluctuations with momenta $-\mathbf{k}_1$ and $-\mathbf{k}_2$, in a region $A$ of the Hubble patch. Graviton B interacts analogously with $\mathbf{k}_3$ and $\mathbf{k}_4$, and $-\mathbf{k}_3$ and $-\mathbf{k}_4$, in a region $B$ that is spatially well separated from $A$. The quantum state of the pair of gravitons is a Bell pair of the form \eqref{eq: Bell state of gravitons}, describing entanglement in their polarization/spin. For certain choices of momenta of the scalar fluctuations $\mathbf{k}_i$ and $-\mathbf{k}_i$ ($i=1,2,3,4$), there is an imprinting of the entanglement of the gravitons in the 8-point scalar correlation function, as described in Sec. \ref{subsec: our observable - 8 point func} and \ref{subsec: CHSH ineq from 8-point}. The physical interpretation of the process is further described in Sec. \ref{subsec: our observable - 8 point func}.}
    \label{fig: schematic rep of process}
\end{figure}

J. S. Bell demonstrates in \cite{Bell} that there is a clear mathematical distinction between any classical, local hidden variable (LHV) theory and a quantum-mechanical (QM) framework exhibiting the non-local phenomenon of entanglement. Over the years, extensive work has explored both theoretical and experimental aspects of Bell’s theorem (e.g. \cite{Clauser_Shimony, CHSH, Gerry, Walls-Milburn}). For the purpose of our work, we are interested in replicating each of the needed elements of the Bell experiment in our inflationary paradigm. These elements are listed below and represented in Fig. \ref{fig: bell experiment scheme}, as done also in \cite{Maldacena_2015}. Together with each element, we give a brief introductory description of its realization in our inflationary setup:

\begin{itemize}
    \item Two separate spatial locations, call them Alice's and Bob's location. In our setup, these are two well-separated spatial points in the same inflationary patch, were the interactions described below occur.
    
    \item An entangled Bell state $\ket{\Psi}$, with components at these two locations. In our setup, we take two gravitons (one at each location) entangled in their polarizations as in state \eqref{eq: Bell state of gravitons}, following the same type of state as in \cite{Tejerina2024}. 

    \item Two possible measurements of some physical observable at each location A (B), whose result is dependent on some local variable $\theta$ ($\phi$), or on some random choice between $\theta$ and $\theta'$ ($\phi$ and $\phi'$). Each observation/measurement is represented by non-commuting operators, call them $A(\theta)$ and $A(\theta')$, and $B(\phi)$ and $B(\phi')$, respectively
    \footnote{This implies that upon two measurements of the same observable (one after the other), the result of the latter measurement is affected by the fact that the former has been previously measured; $[A(\theta),A(\theta')] \neq 0$ and $[B(\phi),B(\phi')] \neq 0$.}. Our physical observable will be a particular momentum configuration of a correlation function of primordial scalar fluctuations that interacted with the gravitons at A and B. This interaction allows for the transfer of polarization information, that in turn carries information from the quantum entangled state of the gravitons. We will consider the polarization structure of a partially-disconnected\footnote{The precise meaning of a \textit{partially-disconnected} diagram is discussed in detail in Sec. \ref{subsec: our observable - 8 point func} and appendix \ref{appendix: 8-point function computation}. In short, it refers to the structure of the momentum-conserving Dirac deltas that appear in the computation of the correlation function at hand, i.e. Eq. \eqref{eqap: relevant delta structure of calc}. This means that we look at the terms in the computation with a specific momentum conservation structure that is not the fully-connected one. Schematically, the fully-connected term has a momentum conserving structure of the type $\delta\left(\sum_{i}^N\mathbf{k}_i\right)$, while the structure of a partially-disconnected term is of the type $\delta\left(\sum_{a}^{n_a}\mathbf{k}_a\right) \delta\left(\sum_{b}^{n_b}\mathbf{k}_b\right)\dots\delta\left(\sum_{c}^{n_c}\mathbf{k}_c\right)$, where the total number of momenta is still the same as the fully-connected piece, i.e. $N=n_a + n_b + \dots + n_c$.} diagram contribution to the full correlation function in Eq. \eqref{eq: observable from 8-point} (see Fig. \ref{fig: schematic rep of process} for a schematic physical representation of the process). The variables $\theta$ and $\phi$ will be angles that rotate the basis in which we are probing the polarization structure of these scalar-graviton interactions. Section~\ref{subsec: density matrix formulation} outlines the details. 

    \item Definite outcomes for quantum measurements of the operators.  In classical Bell-type experiments, measurement outcomes are dichotomic, meaning that the response of the measurements is typically \(A(\theta)=\pm 1\) and \(B(\phi)=\pm 1\), where \(\theta\) and \(\phi\) denote the settings of the measurement devices. In standard laboratory realizations, these settings correspond, for instance, to the orientation angles of anisotropic crystals or polarizers that deflect incoming photons into distinct paths depending on their polarization. Trivially, the combined response function \(A(\theta)B(\phi)\) also yields definite outcomes \(\pm 1\). One can then define $C(\theta,\phi)$ as the average over many realizations of the experiment, as in Eq. \eqref{eq: general definition of C(theta,phi)=<A*B>}. In our construction, \(C(\theta,\phi)\) is defined explicitly at the end of Section~\ref{sec: realization of Bell experiment} (see Eq.~\ref{eq: final construction for C(theta,phi)}). There, the observable is obtained from a sum of correlation functions evaluated for specifically selected momentum configurations of scalar fluctuations. This selection procedure effectively isolates the contributions associated with different graviton polarization states that interact with the scalar modes. This strategy follows the general philosophy of \cite{Tejerina2024}, while extending it through a concrete and systematic prescription for choosing the momenta of the scalar fluctuations.  
    
    \item A channel to classically (non-quantum-mecanical evolution) transmit the results of the measurements to a common location where they can be correlated. In our inflationary setup, this communication channel is provided by the evolution of the relevant scalar curvature perturbations, $\zeta$, after their interaction with the entangled graviton state. These modes exit the Hubble horizon during inflation, where their amplitudes become effectively frozen and undergo a quantum-to-classical transition. After inflation ends, they subsequently re-enter the Hubble radius and can influence later-time observables within the causal patch. One late-time observable can then be used to correlate the outcomes of the quantum measurements, or in this case their imprints on the scalar correlations.
\end{itemize}

Given these elements, we can define the observable:

\begin{equation}
\label{eq: S observable}
    S = C(\theta, \phi) + C(\theta', \phi) + C(\theta, \phi') - C(\theta',\phi') \ ,
\end{equation}

where:
\begin{equation}
\label{eq: general definition of C(theta,phi)=<A*B>}
    C(\theta, \phi)=\expval{A(\theta)\,B(\phi)} \ .
\end{equation}
The joint response $AB$ can only be $\pm 1$, by definition. After averaging over many realizations, one obtains $C(\theta,\phi)$. In our construction, the multiple realizations come the fact that the chosen observable is a statistical average over the observed sky today, since it is built entirely out of correlation functions\footnote{Calculating correlation functions of field fluctuations implies averaging over many realizations of the relevant fluctuations by definition.} of interactions between scalar and entangled tensor modes during the inflationary epoch. The observable $S$ is constructed in a way such that any classical local hidden variable theory (LHV) will yield a value $S^\text{LHV}\leq 2$. However, for an entangled Bell state \cite{Bell}, the value given by the quantum mechanical expectation value can be larger, $S^\text{QM}\leq 2\sqrt{2}$ (see e.g. Ch 9.6 of \cite{Gerry} for a simple proof). By exceeding the former inequality experimentally, one proves non-locality as intrinsic to the quantum-mechanical nature. This particular Bell inequality is called the Clauser-Horne-Shimony-Holt inequality \cite{CHSH}.

\section{Realization of the Bell experiment} \label{sec: realization of Bell experiment}

Let us now introduce the specific computation that will allow us to perform a Bell experiment during inflation. First we introduce in Subsec. \ref{subsec: initial entangled state} and \ref{subsec: density matrix formulation} the quantum entangled state of gravitons that we use throughout the work. Then, we present in Subsec. \ref{subsec: our observable - 8 point func} the observable that we use to record the information about the entangled state (graviton-scalar-scalar interactions, and its physical interpretation. In Subsec. \ref{subsec: CHSH ineq from 8-point} we present the construction of our specific CHSH variable that would allow us to correlate today the imprints on the scalar fluctuations of the non-local correlations from the original entangled state, and get a Bell-inequality violation.

\subsection{Initial state of entangled gravitons} \label{subsec: initial entangled state}

Our initial assumption is the presence of some mechanism that will provide a vacuum state that is not Bunch-Davis, but that it is populated with gravitons entangled by pairs in their polarizations, in the state:

\begin{equation}
    \hspace{-0.3cm} \ket{\Psi}_\text{Bell}=\frac{1}{\sqrt{2}} \Bigl(\ket{+,\mathbf{p}_1}_1 \otimes \ket{+,\mathbf{p}_2}_2 + \ket{\times,\mathbf{p}_1}_1 \otimes \ket{\times,\mathbf{p}_2}_2\Bigr)
    \label{eq: Bell state of gravitons}
\end{equation}
where $\ket{+,\mathbf{p}_1}_1$ represents the eigenstate ``plus'' of the two possible tensor polarizations of graviton 1 (with momentum $\mathbf{p}_1$), and $\ket{\times,\mathbf{p}_1}_1$ represents the other tensor polarization ``cross'' for graviton 1, and analogously for graviton 2. In principle, one can also add the probability of this state occurring, but in later sections we normalize our observable by everything except the polarization terms of the correlator, and hence this factor will be normalized out.\\

This state could potentially be obtained through two cubic-vertex interactions or a single quartic-vertex interaction (two scalars in and two tensors out) (see \cite{Smatrix_dS} for the de Sitter S-matrix formalism). More specifically, in order to get a state like \eqref{eq: Bell state of gravitons}, one would need to evaluate the polarization tensors at opposite momenta which would imply the initial state should have zero total momentum. This could be due to two scalars interacting through a propagator to give two gravitons for instance using cubic vertices, or using a single quartic vertex as mentioned above. An explicit calculation for the 4-vertex is shown in appendix B of \cite{Tejerina2024}. Another perspective is just to postulate the given state, and study the implications it would have on the detectability of quantum effects from inflation\footnote{In principle, one can postulate any choice of entangled 2-state for the system of two gravitons and study the implications. We choose state \eqref{eq: Bell state of gravitons} for concreteness. The analysis of the observable presented in the following sections changes once the probabilities specific to this state are substituted, i.e. from Eq. \eqref{eq: observable after substituting bell state probabilities} onward.}. Mainly, this work is not concerned with how this state would be produced, but rather we postulate it and study the detectability of its ``quantumness'' through the construction of a Bell experiment, like many others have looked for quantum signatures e.g. \cite{Martin:2018lin,Kanno_2019,J.Martin_V.Venin, Maldacena_2015}.

We emphasize that entangled graviton \emph{pairs} provide the leading contribution to the quantum correlations relevant for our analysis. In perturbation theory around the inflationary background, the production of two gravitons arises already at the lowest non-trivial order from cubic interaction vertices. By contrast, processes that generate entangled states involving three or more gravitons require either additional interaction vertices in the corresponding scattering amplitudes, or the inclusion of higher-order terms in the action. 
Both effects lead to a parametric suppression. Diagrams with extra vertices are suppressed by additional powers of the slow-roll parameters and of the interaction scale, while higher-order operators in the effective action are themselves subleading with respect to the lower-order interactions. As a result, multi-particle graviton production -- and the associated higher-order entangled states -- contributes at subleading order in the perturbative expansion. 
Consequently, the dominant quantum correlations generated during inflation are expected to arise from the production of entangled graviton pairs, with contributions from states containing three or more gravitons providing only suppressed corrections.

\vspace{1cm}

\subsection{Density matrix formulation for the Bell state of gravitons} \label{subsec: density matrix formulation}

The purpose of this section is expressing the entangled state in a density matrix form that will make the latter calculations more explicit. We take the entangled state in Eq. \eqref{eq: Bell state of gravitons}:
\begin{align}
    \lvert\Psi\rangle
& =\frac{1}{\sqrt{2}} \Bigl(\ket{+,\mathbf{p}_1}_1 \otimes \ket{+,\mathbf{p}_2}_2 + \ket{\times,\mathbf{p}_1}_1 \otimes \ket{\times,\mathbf{p}_2}_2\Bigr) \notag \\
&\equiv \frac{1}{\sqrt{2}}\Bigl(\lvert+,+\rangle + \lvert\times,\times\rangle\Bigr) \notag \;,
\end{align}

where we have hidden the tensor product and momentum for notational convenience. Its density operator is
\[
\rho \;=\; \lvert\Psi\rangle\langle\Psi\rvert
=\frac{1}{2}
\begin{pmatrix}
1 & 0 & 0 & 1 \\
0 & 0 & 0 & 0 \\
0 & 0 & 0 & 0 \\
1 & 0 & 0 & 1
\end{pmatrix}.
\]

For graviton 1, we rotate the basis elements by angle $\theta$ about the axis which corresponds to its direction of propagation. Note that the physical meaning of this angle is as follows: the momentum $\mathbf{p_1}$ of graviton 1 will be equal to $\mathbf{k_1}+\mathbf{k_2}$ (see \ref{eqap: relevant delta structure of calc}). The angle $\theta$ has the interpretation of being the angle in the plane orthogonal to $\mathbf{k_1}+\mathbf{k_2}$. Note that $\mathbf{k_1}, \mathbf{k_2}$ are momenta of scalar fluctuations that we observe today (see Sec. \ref{subsec: our observable - 8 point func}). Under this rotation, the polarization states transform as
\begin{align*}
&\lvert\,\theta\rangle_{1}
= \cos2\theta\,\lvert +\rangle_{1} \;+\;\sin2\theta\,\lvert \times\rangle_{1}\,\,\,,\\
&\lvert\theta^{\perp}\rangle_{1}
= -\,\sin2\theta\,\lvert +\rangle_{1} \;+\;\cos2\theta\,\lvert \times\rangle_{1}\,\,\,,
\end{align*}
where $\ket{\theta^\perp}$ represents state orthogonal to the state $\ket{\theta}$.\\
Equivalently for graviton 2, rotating the basis elements analogously by an angle~$\phi$ gives
\begin{align*}
&\lvert\,\phi\rangle_{2}
= \cos2\phi\,\lvert +\rangle_{2} \;+\;\sin2\phi\,\lvert \times\rangle_{2}\,\,\,,\\
&\lvert\phi^{\perp}\rangle_{2}
= -\,\sin2\phi\,\lvert +\rangle_{2} \;+\;\cos2\phi\,\lvert \times\rangle_{2}\,\,\,.
\end{align*}

After substituting the rotated eigenstates in the entangled state, and collecting terms one obtains

\begin{widetext}
\begin{equation}
\lvert\Psi\rangle
=\frac{1}{\sqrt2}\Bigl(
\cos(2(\theta-\phi))\,\lvert\theta,\phi\rangle
+\sin(2(\theta-\phi))\,\lvert\theta,\phi^\perp\rangle
-\sin(2(\theta-\phi))\,\lvert\theta^\perp,\phi\rangle
+\cos(2(\theta-\phi))\,\lvert\theta^\perp,\phi^\perp\rangle
\Bigr)
\end{equation}
\end{widetext}

where e.g.\ \(\lvert\theta,\phi\rangle=\lvert\theta\rangle_1\otimes\lvert\phi\rangle_2\).\\

In the ordered basis
\[
\bigl\{\lvert\theta,\phi\rangle,\;\lvert\theta,\phi^\perp\rangle,\;
\lvert\theta^\perp,\phi\rangle,\;\lvert\theta^\perp,\phi^\perp\rangle\bigr\},
\]
the amplitude vector of \(\lvert\Psi\rangle\) is
\[
\ket{\Psi} = \frac{1}{\sqrt2}
\begin{pmatrix}
\cos(2(\theta-\phi)) \\[4pt]
\sin(2(\theta-\phi)) \\[4pt]
-\sin(2(\theta-\phi)) \\[4pt]
\cos(2(\theta-\phi))
\end{pmatrix}.
\]
Thus the density matrix

\[
\rho(\theta,\phi)
=\frac12
\left[\!
\begin{matrix}
c^2 & c\, s & - c\, s & c^2 \\[4pt]
c\, s & s^2 & -s^2 & c\, s \\[4pt] 
-c\, s & - s^2 & s^2 & -c\, s \\[4pt]
c^2 & c\, s & -c\, s & c^2
\end{matrix}
\!\right]\,\,,
\vspace{0.5cm}
\]

where $c\equiv\cos(2(\theta-\phi))$ and $s\equiv\sin(2(\theta-\phi))$.\\

We note that the ``passive viewpoint'' , where we rotate the ordered basis $\{\ket{+},\ket{\times}\}$ into $\{\ket{\theta},\ket{\theta^\perp}\}$, is the usual procedure in the standard setup for a Bell experiment (see e.g. \cite{Gerry}). However, in our cosmological setup computation, it is useful to also write the density matrix in the ``active viewpoint'' i.e. when one rotates the state while keeping the basis the same. The density matrix in the active viewpoint is:

\begin{equation}
\label{eq: active density matrix}
    \rho_{\mathrm{act}}(\theta,\phi)
\;= \;\frac12
\left[\!
\begin{matrix}
c^2 & - c\,s & c\,s & c^2 \\[4pt]
-c\,s & s^2 & -s^2 & -c\,s \\[4pt]
c\,s & -s^2 & s^2& c\,s \\[4pt]
c^2  & -c\,s & c\,s & c^2
\end{matrix}
\!\right] \\
\vspace{0.5cm}
\end{equation}

Working with the active density matrix simplify our calculations throughout the paper. Ultimately, once the trace is taken to compute the expectation value of an operator $\mathcal{O}$, the two formulations are equivalent because of the simple identity proven in appendix \ref{appendix: passive vs active}:

\begin{align}
    & \langle ++|\rho_\text{active}|++\rangle
= \langle \theta,\phi|\rho_{\rm passive}|\theta,\phi\rangle  \notag \\ & \notag 
 \notag
\end{align}

\subsection{Choice of observable: Partially-disconnected 8-point scalar correlation function} \label{subsec: our observable - 8 point func}

In this section, we introduce and interpret the relevant physical process in this work. Through it, we aim to define some observable $O$ corresponding to the expectation value of an operator $\mathcal{O}$ with respect to the quantum state $\Psi$ that we will use to construct the CHSH inequality in Eq. \eqref{eq: S observable}.

The observable $O$ should be some quantity potentially measurable through observation of our Universe. We propose an observable that is potentially measurable through observation of LSS. The observable described in the following computation is constructed exclusively from interactions between scalar and tensor fluctuations of the inflationary fields that replicate the behavior of the standard Bell experiment. This observable comes from computing the expectation value of the operator corresponding to 8 scalar curvature perturbations with opposite pairs of momenta $\mathbf{k}_a$ and $-\mathbf{k}_a$ ($a=1,2,3,4$). This operator is:
\begin{equation}
\label{eq: 8-point function mathcal{O}}
\hspace{-0.3cm}    \mathcal{O}(\mathbf{k}_1,\mathbf{k}_2,\mathbf{k}_3,\mathbf{k}_4) = \zeta_{\mathbf{k}_1} \zeta_{\mathbf{k}_2}\zeta_{\mathbf{k}_3}\zeta_{\mathbf{k}_4}\zeta_{\mathbf{-k}_1}\zeta_{\mathbf{-k}_2}\zeta_{\mathbf{-k}_3}\zeta_{\mathbf{-k}_4}
\end{equation}

The expectation value of $\mathcal{O}$ is computed with respect to the entangled graviton state \eqref{eq: Bell state of gravitons}, also encoded in the density matrix \eqref{eq: active density matrix}.\\

Before diving into the calculation, let us give a physical interpretation, also depicted in Fig. \ref{fig: schematic rep of process}. This process can be viewed as the graviton at location A interacting with two different pairs of scalar fluctuations $\zeta$ with momenta $\mathbf{k}_1,\,\mathbf{k}_2,$ and $-\mathbf{k}_1, \,-\mathbf{k}_2$, respectively.

The other graviton at location B is interacting with another two pairs of scalar fluctuations with momenta $\mathbf{k}_3,\,\mathbf{k}_4,$ and $-\mathbf{k}_3,\, -\mathbf{k}_4$. The interaction vertices couple each of the gravitons to pairs of scalar fluctuations, and information about the shared (entangled) quantum state of the gravitons' polarization can be transferred to the scalars. This transfer is due to the coupling between the spatial derivatives of the scalars and the polarization tensor of the graviton in the scalar-scalar-tensor terms coming from the vertex corresponding to action \eqref{eq: graviton scalar scalar action}.

Concretely, certain selections of momenta will give specific imprints of the polarization of the gravitons on the scalar fluctuation correlations. This will allow us to imprint and measure, in the joint response of all the scalar fluctuations (the 8-point function), the non-local correlations implicit in the entangled state of gravitons.\\

Let us sketch some features of the calculation:
\begin{itemize}
    \item We will be computing the expectation value of a scalar 8-point function with respect to a state of inflation populated with pairs of gravitons entangled in their polarizations. 
    \item We do not compute the full correlation function, but rather the contribution from a specific momentum conservation structure; we already referred to it as a partially-disconnected contribution in the previous sections. The details of the structure are presented in appendix \ref{appendix: 8-point function computation}. 
    \item The 8-point function will reduce to a pair of 2-point correlations with momenta $\mathbf{k}_a$, and their two mirror correlations $-\mathbf{k}_a$, with $a=1,2,3,4$ (because of the structure of the particular choice of this contribution, see appendix \ref{appendix: 8-point function computation}). 
    \item We will be using the \textit{in-in} formalism. Here, these correlations are given by four interaction vertices of scalar-scalar-graviton ($\zeta\zeta\gamma$, see action in Eq. \ref{eq: graviton scalar scalar action}). This gives a total of eight scalars, and ``four'' gravitons (two for the \textit{in}-ket state, and two for the \textit{in}-bra), as depicted in Fig. \ref{fig: schematic rep of process}. We note that there are only two gravitons under consideration, and that the presence of ``four'' is merely a feature of the \textit{in-in} formalism.
\end{itemize}

Given the operator $\mathcal{O}$ in \eqref{eq: 8-point function mathcal{O}}, our observable is:
\begin{widetext}
    \begin{align}
    & \hspace{-1.1cm}  O(\theta,\phi,\mathbf{k}_1,\mathbf{k}_2,\mathbf{k}_3,\mathbf{k}_4) \equiv \expval{\zeta_{\mathbf{k}_1} \zeta_{\mathbf{k}_2}\zeta_{\mathbf{k}_3}\zeta_{\mathbf{k}_4}\zeta_{\mathbf{-k}_1}\zeta_{\mathbf{-k}_2}\zeta_{\mathbf{-k}_3}\zeta_{\mathbf{-k}_4}}_\rho \notag \\
    & \notag \\
    & \hspace{-1.1cm} = \text{Tr}\Big(\rho_\text{act}\,(\eta, \theta,\phi)\,\zeta_{\mathbf{k}_1} \zeta_{\mathbf{k}_2}\zeta_{\mathbf{k}_3}\zeta_{\mathbf{k}_4}\zeta_{\mathbf{-k}_1}\zeta_{\mathbf{-k}_2}\zeta_{\mathbf{-k}_3}\zeta_{\mathbf{-k}_4}\Big) \notag \\ & \notag \\
    & \hspace{-1.1cm} = \frac{\text{Tr}\Big(\rho_\text{act}\,(\theta,\phi)\, U_{-\infty}(\eta) \zeta_{\mathbf{k}_1} \zeta_{\mathbf{k}_2}\zeta_{\mathbf{k}_3}\zeta_{\mathbf{k}_4}\zeta_{\mathbf{-k}_1}\zeta_{\mathbf{-k}_2}\zeta_{\mathbf{-k}_3}\zeta_{\mathbf{-k}_4} U_{-\infty}(\eta) \Big)}{\text{Tr}\Big(\rho_\text{act}(\theta,\phi) U_{-\infty}(\eta) _{-\infty}(\eta) \Big)} \notag \\  & \notag \\   
    & \hspace{-1.1cm} \left. = \frac{\sum_a \, p_a\ \left._\text{in}\bra{\psi_a}\right. U_{-\infty}(\eta)\left(\prod_{n=1}^8\zeta_{\mathbf{k}_n} \right) U_{-\infty}(\eta)  \ket{\psi_a}_\text{in}}{\text{Tr}\Big(\rho_\text{act}(\theta,\phi) \Big)} \right|_{\mathbf{k}_{j} = -\mathbf{k}_{i}} \label{eq: observable from 8-point} 
\end{align}
\end{widetext}
where $i=1,\dots,4$ and $j=5,\dots,8$. See chapter 1 of \cite{kamenev2023field} for reference. Note that $U_{-\infty}(\eta)\equiv U(\eta, -\infty)$ is the time evolution operator. In our case we will set the conformal time $\eta =0$ since we want to evaluate the correlators at the end of inflation. Note that in the in-in formalism, the expectation value at time $\eta$ already includes the full quantum evolution, from the initial state in the far past, to the final time $\eta$ through the time-evolution operators, thus the graviton-scalar interactions are accumulated up to the final time $\eta$. We choose $\eta=0$ as a convenient value for the final time which corresponds to the end of inflation because by that time all the relevant modes have become superhorizon and the correlator has already frozen by then. 

We also interpret the last line as: 
\begin{enumerate}
    \item We compute the quantum expectation value of the 8-point scalar correlation function, dependent on all momenta $\mathbf{k}_1,\dots,\mathbf{k}_8$.
    
    \item We evaluate the result at the momenta $\mathbf{k}_5=~-\mathbf{k}_1, \,\mathbf{k}_6=-\mathbf{k}_2, \,\mathbf{k}_7=-\mathbf{k}_3,\,\mathbf{k}_8=-\mathbf{k}_4$.
\end{enumerate}
Here, $\rho_\text{act}$ encodes the (rotated) gravitons' entangled state of our inflationary setup; the probabilities $p_a$ are the diagonal terms in $\rho_\text{act}$. The $\ket{\psi_a}$ correspond to the eigenstates of the system (in our case $\ket{++}$, $\ket{+\times}$, $\ket{\times+}$ and $\ket{\times\times}$, or their rotated versions $\lvert\theta,\phi\rangle, \,\lvert\theta,\phi^\perp\rangle,\,
\lvert\theta^\perp,\phi\rangle$ and $ \lvert\theta^\perp,\phi^\perp\rangle$). \\

The time evolution operator can be written as:

\begin{equation}
 U(\eta, \eta_0)
= \mathcal{T} \exp\!\Biggl(-\,i \int_{\eta_0}^{\eta} d\eta'\; H_I(\eta')\Biggr)
\end{equation}

\begin{equation}
 U(\eta_0, \eta)
= \tilde{\mathcal{T}} \exp\!\Biggl(+\,i \int_{\eta_0}^{\eta} d\eta'\; H_I(\eta')\Biggr)
\end{equation}

where $\mathcal{T}$ denotes the time order. Note that we have not included the so called epsilon-prescription which is required to adiabatically turn off the interactions in the asymptotic past. The reason is that we will not explicitly use this prescription in our calculations because we do not explicitly evaluate the time integrals. All that is needed is the polarization structure part of the correlation function; hence it is important to note that the epsilon-prescription is still there, but since we made no use of it, we omit it from our formulas for notational simplicity.\\

Let us first calculate the general expectation value with respect to an initial state populated with two gravitons. The density matrix will later specify the probabilities given by the entangled state. The general calculation is performed in the \textit{in-in} formalism. \\

The relevant scalar-scalar-tensor vertex is computed in \cite{Maldacena:2002}. The corresponding third-order action in the spatially flat gauge \eqref{eq: spatially flat gauge} is:

\begin{equation}
   I^{(3)} \supset \frac{1}{2}\int \text{d}^4x\, \epsilon \, \widetilde{\gamma}_{ij}\, \partial^{\,i}\varphi \, \partial^{\,j}\varphi + O(\epsilon^2)
\end{equation}
where $\widetilde{\gamma}_{ij}$ carries the tensor degree of freedom, and becomes $\gamma_{ij}$ by field redefinitions (which vanish outside the horizon). In this gauge, the inflaton $\varphi$ (at Hubble crossing) is related to the scalar curvature perturbation \cite{Maldacena:2002, Seery_2009_Trispec_GE} by:
\begin{equation}
    \zeta = -\frac{H}{\dot{\phi}}\,\varphi = - \frac{\varphi_*}{\sqrt{2\epsilon}}
\end{equation}
where the $*$ denotes evaluation at the time of horizon crossing. Thus, we can work with the action written in the following form:
\begin{equation}
   I^{(3)} \supset \int \text{d}\eta\,\text{d}^3\mathbf{x}\, \epsilon \, a^2 \gamma^{ij} \, \partial_i\zeta \, \partial_j\zeta 
   \label{eq: graviton scalar scalar action}
\end{equation}
that is first order in $\epsilon$. In this gauge, it is the only term for this type of vertex, which simplifies enormously the computation. Also, it allows us to work directly with the physical degree of freedom $\zeta$.\\

To perform the computation in the \textit{in-in} formalism, we first note that any general \textit{in}-states can be written as a linear combination of elements of the basis of the form:

\begin{equation}
        \ket{\psi_a}_\text{in} = \ket{s_1,\mathbf{p}_1}_1 \otimes \ket{s_2,\mathbf{p}_2}_2 = {b_{\mathbf{p}_1}^{s_1}}^\dagger\,{b_{\mathbf{p}_2}^{s_2}}^\dagger \ket{0}_\text{in}
        \vspace{0.2cm}
\end{equation}
where $b_{\mathbf{p}_i}^{s_i}$ and ${b_{\mathbf{p}_i}^{s_i}}^\dagger$ are annihilation and creation operators of a graviton (spin-2 excitation) with momentum $\mathbf{p}_i$ and polarization/spin $s_i = +, \times$, i.e.
\begin{equation}
    b_{\mathbf{p}_i}^{s_i}\ket{0}_\text{in} = 0 \,\,\,\,\,(i=1,2) \text{  } \forall \,\mathbf{p}
\end{equation}

Here $\ket{0}_\text{in}$ is the Bunch Davies vacuum.\\

Ignoring for now the normalization term, each term in the last line of our observable \eqref{eq: observable from 8-point} can be computed as:
\begin{widetext}
    \begin{align}
\label{eq: 8-point exp val from braket}
& \left._\text{in}\bra{\psi_a}\right. U(\eta,-\infty)\left(\prod_{n=1}^8\zeta_{\mathbf{k}_n} \right) U(-\infty,\eta)  \ket{\psi_a}_\text{in} \notag \\
 & =  \bra{0} b_{\mathbf{p}_1}^{s_1}b_{\mathbf{p}_2}^{s_2} \,\Tilde{\mathcal{T}}\left\{ e^{i \int_{-\infty}^\eta H_I \text{d}\eta'}\right\}  \, \left(\prod_{n=1}^8\zeta_{\mathbf{k}_n} \right) \,  \, \mathcal{T} \left\{ e^{-i \int_{-\infty}^\eta H_I \text{d}\eta'}\right\} \left(b_{\mathbf{p}_1}^{s_1}\right)^\dagger \left(b_{\mathbf{p}_2}^{s_2}\right)^\dagger \ket{0} 
\end{align}
\end{widetext}

where the interaction Hamiltonian $H_I$ is given by the action in Eq. \eqref{eq: graviton scalar scalar action}. The full computation is shown in Appendix \ref{appendix: 8-point function computation}. In the calculation, we take as the term of interest one of the contractions corresponding to a specific momentum conservation structure. This term corresponds to a partially-disconnected  diagram with four 3-vertex interactions of the type given by the action \eqref{eq: graviton scalar scalar action}. The partially disconnected structure of the correlator can be seen as a consequence of the cluster decomposition property \cite{weinberg1995quantum} where the correlator will split into 2 (or more) pieces as a consequence of having parts of the system being spatially separated and effectively non-interacting. After manipulations, and factorizing the terms with polarization information, we obtain for the expectation value of our operator a term of the form

\begin{align}
    \sim & F(\mathbf{k}_1,\mathbf{k}_2,\mathbf{k}_3,\mathbf{k}_4,\mathbf{k}_5,\mathbf{k}_6,\mathbf{k}_7,\mathbf{k}_8)  \times \notag \\
    & \epsilon_{ij}^{s_1}(\mathbf{k}_{12}) k_1^i k_2^j  \, \epsilon_{lm}^{s_1}(\mathbf{-k}_{56}) k_5^l k_6^m \, \epsilon_{pq}^{s_2}(\mathbf{k}_{34}) k_3^p k_4^q  \, \epsilon_{rt}^{s_2}(\mathbf{-k}_{78}) k_7^r k_8^t 
\end{align}

where $\mathbf{k}_{ab}\equiv \mathbf{k}_a+\mathbf{k}_b$. For simplicity, we absorb the normalization factor in \eqref{eq: observable from 8-point} into $F$ without loss of generality.

We now select four out of the eight momenta $\mathbf{k}_5 = -\mathbf{k}_1$, $\mathbf{k}_6 = -\mathbf{k}_2$, $\mathbf{k}_7 = -\mathbf{k}_3$, $\mathbf{k}_8 = -\mathbf{k}_4$, and get:

\begin{align}
\label{eq: polariz tensors term}
&
\left._\text{in}\bra{\psi_a}\right. U(0,-\infty)\left(\prod_{n=1}^8\zeta_{\mathbf{k}_n} \right) U(-\infty, 0)  \ket{\psi_a}_\text{in} \notag \\   
& = F(\mathbf{k}_1,\mathbf{k}_2,\mathbf{k}_3,\mathbf{k}_4,\mathbf{-k}_1,\mathbf{-k}_2,\mathbf{-k}_3,\mathbf{-k}_4) \notag \\
& \quad\quad\quad\quad \times \big(\epsilon_{ij}^{s_1}(\mathbf{k}_{12}) k_1^i k_2^j \, \epsilon_{lm}^{s_2}(\mathbf{k}_{34}) k_3^l k_4^m \big)^2\,\,,
\end{align}

where we have now made explicit that we are evaluating the expression at the end of inflation where $\eta = 0$. The term in \eqref{eq: polariz tensors term} that contains the polarization tensors will be the key in our construction of the CHSH variable that will yield a Bell violation.\\

If we track down the powers of slow-roll parameter $\epsilon$ and the primordial Hubble parameter $H$, we find that they would appear as a prefactor in the $F$ factor like:
\begin{equation}
    F\sim \frac{H^8}{\epsilon^4} \sim P_\zeta^4
\end{equation}

where $P_\zeta$ is the well-known primordial scalar power spectrum.
\vspace{1cm}

\subsection{CHSH inequality from the 8-point correlation function} \label{subsec: CHSH ineq from 8-point}

Having defined our cosmological observable coming from a specific contribution to the scalar 8-point function, we now make contact with the density matrix formulation, where $\rho_\text{act}$ in Eq. \eqref{eq: active density matrix} represents the quantum state of our initial state of gravitons entangled in their polarizations. Putting together the elements that we have previously discussed, we construct a CHSH variable for a Bell test. \\

We recall that the above calculation is a valid result for any polarization of the initial graviton state. For our specific entangled (rotated\footnote{Recall that we have rotated the state with respect to two different angles $\theta,\,\phi$ corresponding to the measure at Alice's and Bob's location, respectively. This is why we have all four components of ``plus'' and ``cross'' polarization, instead of the initial two components $++$ and $\times \times$.}) state, the probabilities of the state are all encoded in the diagonal terms of the density matrix. Going back to \eqref{eq: observable from 8-point} and using our result for \eqref{eq: polariz tensors term}, our observable is:

\begin{widetext}
    \begin{align}
     & O(\theta,\phi,\mathbf{k}_1,\mathbf{k}_2,\mathbf{k}_3,\mathbf{k}_4) \equiv \text{Tr}\Big(\rho(\theta,\phi)\,\zeta_{\mathbf{k}_1} \zeta_{\mathbf{k}_2}\zeta_{\mathbf{k}_3}\zeta_{\mathbf{k}_4}\zeta_{\mathbf{-k}_1}\zeta_{\mathbf{-k}_2}\zeta_{\mathbf{-k}_3}\zeta_{\mathbf{-k}_4}\Big)  = \sum_a \, p_a\, \left._\text{in}\bra{\psi_a}\right. U(0,-\infty)\left(\prod_{n=1}^8\zeta_{\mathbf{k}_n} \right) U(-\infty,0)  \ket{\psi_a}_\text{in}  \notag\\ & \notag \\
     & \hspace{2cm}= F(\mathbf{k}_1,\mathbf{k}_2,\mathbf{k}_3,\mathbf{k}_4,-\mathbf{k}_1,-\mathbf{k}_2,-\mathbf{k}_3,-\mathbf{k}_4)  \notag \\
     &\hspace{2cm}\times \frac{1}{2} \Bigg( \big(\epsilon_{ij}^{+}(\mathbf{k}_{12}) k_1^i k_2^j \, \epsilon_{lm}^{+}(\mathbf{k}_{34}) k_3^l k_4^m\big)^2\cos^2(2(\theta-\phi)) + \big(\epsilon_{ij}^{+}(\mathbf{k}_{12}) k_1^i k_2^j \, \epsilon_{lm}^{\times}(\mathbf{k}_{34}) k_3^l k_4^m\big)^2\sin^2(2(\theta-\phi)) \notag \\
     &\hspace{2cm} +   \big(\epsilon_{ij}^{\times}(\mathbf{k}_{12}) k_1^i k_2^j \, \epsilon_{lm}^{+}(\mathbf{k}_{34}) k_3^l k_4^m\big)^2\sin^2(2(\theta-\phi))  + \big(\epsilon_{ij}^{\times}(\mathbf{k}_{12}) k_1^i k_2^j \, \epsilon_{lm}^{\times}(\mathbf{k}_{34}) k_3^l k_4^m\big)^2\cos^2(2(\theta-\phi)) \Bigg)
     \label{eq: observable after substituting bell state probabilities}
\end{align}
\end{widetext}

where in the equality of the second line we substitute expression \eqref{eq: polariz tensors term}, together with the various probabilities $p_a$ encoded in $\rho(\theta,\phi)$ for the eigenstates of the system in the basis $$\bigl\{ 
\ket{\psi_a(\theta,\phi)} \bigr\}\equiv~ \bigl\{\lvert\theta,\phi\rangle,\;\lvert\theta,\phi^\perp\rangle,\;
\lvert\theta^\perp,\phi\rangle,\;\lvert\theta^\perp,\phi^\perp\rangle\bigr\}.$$

This observable naturally encodes all the probabilities at various angles. \\

In order to perform a Bell experiment one needs the operator sensitive to the entangling variable (i.e. the local ``measuring device'') to yield different results if the single incoming entangled particle (in our case the each graviton) at one location was in $+$ or $\times$. In an example of a standard realization of a Bell experiment, this ``local measuring device'' is a calcite crystal that separates the incoming single state (photon ray) into two different ones with different photon polarizations (o-ray for horizontal polarization and e-ray for vertical polarization). The different responses of the detector are $A=+1$ for the o-ray and $A=-1$ for the e-ray (idem for B) (see e.g. Sec 9.6 of \cite{Gerry}). Therefore, we want the ``local measuring device'' of our gravitons to yield a collective response $AB$ of $+1$ for $A=1,B=1$ ($++$ term) and $A=-1,B=-1$ ($\times \times$ term), and of $-1$ for $A=1,B=-1$ ($+\times$ term) and $A=-1,B=1$ ($\times +$ term). Each response will have a certain probability depending on the quantum state under evaluation. 

In our setup, the ``local measuring device'' at each location is the 3-point vertex interaction $\gamma\zeta\zeta$, whose interactions or ``results of the measure'' upon horizon crossing of the gravitons (which is the interaction that we are interested in) we recorded in the observable of Eq. \eqref{eq: observable after substituting bell state probabilities}. To model the same kind of behavior for the ``measuring device'' as standard Bell experiments (to model the calcite crystal), we show that by choosing the various configurations of the momenta $\mathbf{k}_a$, we can isolate each of the $++$, $+\times$, $\times+$ and $\times \times$ terms, as described in the following discussion.\\

First, we normalize $O$ by the $F$ factor i.e. $O \to \frac{O}{F}$ and obtain:

\begin{align}
\label{eq: normalized O}
    O (\theta,\phi,&\mathbf{k}_1,\mathbf{k}_2,\mathbf{k}_3,\mathbf{k}_4) =  \notag \\
     & = \frac{1}{F}\,\expval{\zeta_{\mathbf{k}_1} \zeta_{\mathbf{k}_2}\zeta_{\mathbf{k}_3}\zeta_{\mathbf{k}_4}\zeta_{\mathbf{-k}_1}\zeta_{\mathbf{-k}_2}\zeta_{\mathbf{-k}_3}\zeta_{\mathbf{-k}_4}}_\rho  \notag \\ & \notag \\
     & = \frac{1}{2}\Bigl(P_+^{(A)} P_+^{(B)}\Bigr)^2\cos^2(2(\theta-\phi)) \notag \\ & \notag \\
     & + \frac{1}{2}\Bigl(P_+^{(A)} P_\times^{(B)}\Bigr)^2\sin^2(2(\theta-\phi)) \notag \\ & \notag \\
     & + \frac{1}{2}\Bigl(P_\times^{(A)} P_+^{(B)}\Bigr)^2\sin^2(2(\theta-\phi)) \notag \\ & \notag \\
     & + \frac{1}{2}\Bigl(P_\times^{(A)} P_\times^{(B)} \Bigr)^2\cos^2(2(\theta-\phi)) 
\end{align}
where we defined:
\begin{equation*}
        P_{s_1}^{(A)}\equiv P_{s_1}^{(A)}(\mathbf{k}_1, \mathbf{k}_2) = \epsilon_{ij}^{s_1}(\mathbf{k}_{12}) k_1^i k_2^j 
\end{equation*}
\begin{equation*}
    P_{s_2}^{(B)} \equiv P_{s_2}^{(b)}(\mathbf{k}_3, \mathbf{k}_4) = \epsilon_{lm}^{s_2}(\mathbf{k}_{34}) k_3^l k_4^m
\end{equation*}
The quadratic sine and cosine terms in \eqref{eq: normalized O} are the probabilities of the vertex interactions at separate locations with a collective response of $++$, $+\times$, $\times+$ and $\times \times$ (after the individual states have been rotated by angles $\theta, \phi$). 

Now, to isolate these terms that contain the different probabilities, we express our vectors in spherical coordinates, as done also in \cite{Seery_2009_Trispec_GE}. For location A (everything works analogously in B), we choose the sum $\mathbf{k}_{12}$ in the $\hat{\mathbf{z}}$-direction, and choose two orthonormal vectors $\mathbf{e}_{12},\bar{\mathbf{e}}_{12}$. We express the polarization term of the graviton at this location in the basis $\{\mathbf{k}_{12}, \mathbf{e}_{12},\bar{\mathbf{e}}_{12}\}$. Written in matrix form, we have:
\begin{align}
    P_+^{(A)} & \equiv \mathbf{k}_1^T\cdot \epsilon^+(\mathbf{k}_{12})\cdot \mathbf{k}_{2} \notag \\
    & =\frac{1}{\sqrt{2}} \begin{pmatrix}
        k_{1x} & k_{1y} & k_{1z}
    \end{pmatrix}
    \begin{pmatrix}
        1 & 0 & 0 \\
         0 & -1 & 0 \\
        0 & 0 & 0 \end{pmatrix} 
        \begin{pmatrix}
            k_{2x} \\ k_{2y} \\ k_{2z}
        \end{pmatrix} \notag \\
        & = \frac{1}{\sqrt{2}}\,\left(k_{1x}k_{2x} - k_{1y}k_{2y}\right) \notag
\end{align}
\begin{align}
    P_\times^{(A)} & \equiv \mathbf{k}_1^T\cdot \epsilon^\times(\mathbf{k}_{12})\cdot \mathbf{k}_{2} \notag  \\ 
    & = \frac{1}{\sqrt{2}} \begin{pmatrix}
        k_{1x} & k_{1y} & k_{1z}
    \end{pmatrix}
    \begin{pmatrix}
        0 & 1 & 0 \\
         1 & 0 & 0 \\
        0 & 0 & 0 \end{pmatrix} 
        \begin{pmatrix}
            k_{2x} \\ k_{2y} \\ k_{2z}
        \end{pmatrix} \notag \\ 
        & = \frac{1}{\sqrt{2}}\,\left(k_{2x}k_{1y} + k_{1x}k_{2y}\right) \notag
\end{align}

In spherical coordinates\footnote{Note that the angles from the spherical coordinates $\theta_1$, $\phi_1$ should not be confused with the quantum-state rotation angles $\theta$ and $\phi$.}, the basis $\{\mathbf{k}_{12}, \mathbf{e}_{12},\bar{\mathbf{e}}_{12}\}$ has the following properties:
\begin{equation}
\label{eq: properties basis}
    k_{1}\sin\theta_{1} = k_{2}\sin\theta_{2}\,\,\,;\,\,\,\phi_{2} = \phi_{1} + \pi
\end{equation}
where $\theta_i$ is the polar angle with respect to the z-axis$||\mathbf{k}_{12}$, and $\phi_i$ the azimuthal angle, of the vector $\mathbf{k}_i$ in this basis.

In these coordinates:
\begin{align}
    P_+^{(A)} & = k_1 k_2 \sin\theta_1 \sin\theta_2 \left(\cos\phi_1 \cos \phi_2 - \sin\phi_1\sin\phi_2\right) \notag \\
  & = -k_1^2 \sin^2\theta_1 \cos(2\phi_1) \\
  \notag \\
  P_\times^{(A)} & =  k_1 k_2 \sin\theta_1 \sin\theta_2 \left( \sin\phi_1 
 \cos\phi_2 + \cos\phi_1 \sin\phi_2\right) \notag \\
 & = -k_1^2 \sin^2\theta_1 \sin(2\phi_1)
\end{align}

where we made use of properties \eqref{eq: properties basis}. \\

So, for example, selecting the momenta as $k_1,k_2=~1$, $\theta_1=\pi/4$ and $\phi_1=0$ isolates the ``$+$'' polarization component at location A:
\begin{equation}
    P_+^{(A)}=-\sin^2\theta_1 = - \frac{1}{2}\,\,\,;\,\,\,P_\times^{(A)} = 0
\end{equation}

An analogous reasoning for the selection of $+$ and $\times$ of Bob's graviton works for ${\mathbf{k}}_{3},{\mathbf{k}}_4$ in the basis  $\{\mathbf{k}_{34}, \mathbf{e}_{34},\bar{\mathbf{e}}_{34}\}$.\\

We can now isolate specific polarization terms by evaluating our operator $O$ in \eqref{eq: normalized O} for different choice of momenta, as follows:

\begin{align}
\label{eq: O++ for choice of momenta}
  & O_{++}(\theta,\phi) \notag \\
  & \equiv\frac{1}{N}\,
    O\left(\theta,\phi, k_a=1, \theta_1=\theta_3=\frac{\pi}{4} ,\phi_1=\phi_3=0
     \right) \notag\\ 
    & = \frac{1}{2}\cos^2(2(\theta-\phi)) \notag \\
    & = \Pr\bigl(\lvert\theta,\phi\rangle\bigr)
\end{align}

where for the specific choices of angles above we have $1/N=8$. Note that the selection of polarization by choosing a specific configuration of momenta is controlled by the polar angles $\phi_1, \phi_3$, and that the azimuthal angles $\theta_1,\theta_3$ just give an overall normalization factor $N(\theta_1,\theta_3) = \sin^2\theta_1 \sin^2\theta_3$.

This normalization factor imposes the condition that $\theta_1,\theta_3\neq 0$; this is natural, as for this choice all terms in \eqref{eq: normalized O} would vanish. Another constraint is $\theta_1,\theta_3\neq \pi/2$. This is just an effect given by the choice of basis; the normalized vector $\mathbf{k}_{12} || \hat{\mathbf{z}}$ for $\theta_1=\pi/2$ would take a value $\mathbf{k}_{12}=\mathbf{0}$, which has no direction and cannot act as a proper basis (idem for $\mathbf{k}_{34}$).\\

Similar definitions to \eqref{eq: O++ for choice of momenta} lead to the four following observables, which naturally encode the various probabilities (see Table \ref{tab:polar-angles} for the particular choices of momenta in spherical coordinates):

\begin{table*}[t]
\caption{Specific choice of polar angles for selecting polarizations in the
definitions of the observables (\ref{eq: O++}--\ref{eq: Oxx}) from the general
observable \eqref{eq: normalized O}, proportional to the expectation value of
the 8-point scalar correlation function. Note that this is not the only valid
choice of $\phi_1,\phi_3$. The norm of the momenta $\mathbf{k}_a$ can be chosen
to unity ($k_1=k_2=k_3=k_4=1$), and the azimuthal angles to
$\theta_1=\theta_2=\theta_3=\theta_4=\pi/4$, or any
$\theta_a\in(0,\pi/2)$.}
\label{tab:polar-angles}
\begin{ruledtabular}
\begin{tabular}{lcccc}
 & 
 \begin{tabular}[c]{@{}c@{}}
 {\scriptsize$\bigl(P_\times^{(A)}=P_\times^{(B)}=0\bigr)$}\\[0.5ex]
 $O_{++}$
 \end{tabular}
 &
 \begin{tabular}[c]{@{}c@{}}
 {\scriptsize$\bigl(P_\times^{(A)}=P_+^{(B)}=0\bigr)$}\\[0.5ex]
 $O_{+\times}$
 \end{tabular}
 &
 \begin{tabular}[c]{@{}c@{}}
 {\scriptsize$\bigl(P_+^{(A)}=P_\times^{(B)}=0\bigr)$}\\[0.5ex]
 $O_{\times +}$
 \end{tabular}
 &
 \begin{tabular}[c]{@{}c@{}}
 {\scriptsize$\bigl(P_+^{(A)}=P_+^{(B)}=0\bigr)$}\\[0.5ex]
 $O_{\times\times}$
 \end{tabular}
 \\

\hline
$\phi_1$ & 0 & 0 & $\pi/4$ & $\pi/4$ \\
$\phi_3$ & 0 & $\pi/4$ & 0 & $\pi/4$
\end{tabular}
\end{ruledtabular}
\end{table*}

\begin{align}
  O_{++}(\theta,\phi) &= \frac{1}{2}\cos^2(2(\theta-\phi)) =& \Pr\bigl(\lvert\theta,\phi\rangle\bigr) \label{eq: O++}\\
  O_{+\times}(\theta,\phi) &= \frac{1}{2}\sin^2(2(\theta-\phi)) =& \Pr\bigl(\lvert\theta,\phi^\perp\rangle\bigr) \label{eq: O+x} \\
  O_{\times+}(\theta,\phi) &= \frac{1}{2}\sin^2(2(\theta-\phi)) =& \Pr\bigl(\lvert\theta^\perp,\phi\rangle\bigr) \label{eq: Ox+} \\
  O_{\times\times}(\theta,\phi) &= \frac{1}{2}\cos^2(2(\theta-\phi)) =& \Pr\bigl(\lvert\theta^\perp,\phi^\perp\rangle\bigr) \label{eq: Oxx}
\end{align}

Now we can write down the Bell observable precisely as in \cite{Gerry}:

\begin{align}
  C&(\theta,\phi) =  \notag \\
    =\; &\Pr\bigl(\lvert\theta,\phi\rangle\bigr)
    - \Pr\bigl(\lvert\theta,\phi^\perp\rangle\bigr)
    - \Pr\bigl(\lvert\theta^\perp,\phi\rangle\bigr)
    +\Pr\bigl(\lvert\theta^\perp,\phi^\perp\rangle\bigr) \notag \\
    =\; &O_{++}(\theta,\phi) 
         - O_{+\times}(\theta,\phi)
         - O_{\times+}(\theta,\phi)\, 
         + O_{\times\times}(\theta,\phi) \notag \\
         =\; & \cos(4(\theta - \phi))  \label{eq: final construction for C(theta,phi)}
\end{align}

Finally, we have built the Bell observable entirely in terms of the 8-point function observable. For each pair of fixed angles $\theta,\phi$, one would look at the 8-point function at four different configurations (e.g. the choices in Table \ref{tab:polar-angles}). Each of these configurations encodes the probabilities of the possible eigenstates. If we set $\theta =0$, $\theta' =~\pi/8$, $\phi =~\pi/16$, $\phi'=~-\pi/16$ we expect a maximal Bell violation of the CHSH inequality defined in \eqref{eq: S observable}:

\begin{equation}
    S = C(\theta,\phi) + C(\theta',\phi)+C(\theta,\phi')-C(\theta',\phi')  = 2\sqrt2
    \label{eq: S evaluated=2sqrt(2)}
\end{equation}

Meanwhile classically such a thing never occurs, as proven for the CHSH inequality \cite{CHSH} for any general hidden variable theories with $A(\theta,\lambda)$ and $B(\phi,\lambda)$, being $\lambda$ one or multiple classical hidden variables (see e.g. Ch. 9.6 of \cite{Gerry} for a simple proof).

So the full prescription presented is straightforward: 

\begin{enumerate}
    \item The observable is defined in Eq. \eqref{eq: normalized O}, coming from the contribution to the correlation function proportional to the specific contractions of internal and external lines detailed in \eqref{eq: appendix B relevant contribution of correlation}. Note that, since the observable is a correlation function, the statistical average needed in the general definition of $C(\theta,\phi)$ in Eq. \eqref{eq: general definition of C(theta,phi)=<A*B>} is already present in the definition of the observable itself.
    
    \item The quantity $C$ in Eq. \eqref{eq: final construction for C(theta,phi)} is then built entirely out of the observable measured at various momenta configurations for $\mathbf{k}_a$ with $a=1,\dots 4$. Each of the terms (\ref{eq: O++}--\ref{eq: Oxx}) in the definition of $C$ comes from one choice of momenta. Our specific choices are presented in Table \ref{tab:polar-angles} for spherical coordinates in two different basis: $\{\mathbf{k}_{12}, \mathbf{e}_{12},\bar{\mathbf{e}}_{12}\}$ for the pair $\mathbf{k}_{1},\,\mathbf{k}_{2}$ and $\{\mathbf{k}_{34}, \mathbf{e}_{34},\bar{\mathbf{e}}_{34}\}$ for the pair $\mathbf{k}_{3},\,\mathbf{k}_{4}$. These basis are explicitly defined in the previous page.
    
    \item The angles $\theta$ and $\phi$ are defined in the orthogonal planes to $\mathbf{k}_{12}$ and $\mathbf{k}_{34}$, respectively. For the fixed choices of momenta (which set a choice for the orthogonal planes) of Table \ref{tab:polar-angles}, the angles $\theta$, $\theta'$, $\phi$ and $\phi'$ are chosen. The terms  (\ref{eq: O++}--\ref{eq: Oxx}) are evaluated at these angles and $C$ is constructed. For the choices of these angles given in the text, one obtains a value of $S^\text{QM}=2\sqrt{2}>2$ in Eq. \eqref{eq: S evaluated=2sqrt(2)}, which violates the classical Bell-inequality.
\end{enumerate}

\section{Conclusion and future work} \label{sec: Conclusion}

In this work, we present a setup in which one could obtain a Bell-inequality violation from the quantum fluctuations coming from a minimal, slow-roll, single-field inflationary phase. This violation, if experimentally measured, would serve as proof of the true quantum nature of these fluctuations as the primordial seeds for the formation of structures in our Universe.\\

We present a theoretical setup in which all necessary elements of a Bell experiment can be constructed out of the basic elements of the prescribed inflationary phase. Specifically, we start from the assumption of the presence of pairs of gravitons (tensor fluctuations of the metric) entangled in their polarization states. The wavelength of these tensor fluctuations gets stretched by the fast quasi-exponential expansion of inflation. Upon horizon crossing, their amplitude freezes and remains constant until the end of inflation. If we look at third-order interactions, we find interactions between each graviton with scalar curvature fluctuations. In particular, we look at interactions of each graviton, with positions at two different locations, with two different pairs of scalar fluctuations. These interactions are dependent on the polarization of each graviton. This polarization in turn carries quantum (non-local) correlations because it is part of an entangled pair. In total, we have eight scalar fluctuations interacting with two gravitons; this has a contribution to primordial non-Gaussianity at the level of the 8-point primordial correlation function.

We then select a specific momentum-conservation structure (which corresponds to a partial contribution to the full signal). By looking at interacting scalar pairs $\mathbf{k}_1,\, \mathbf{k}_2$ and $-\mathbf{k}_1,\, -\mathbf{k}_2$ for one graviton, and $\mathbf{k}_3,\, \mathbf{k}_4$ and $-\mathbf{k}_3,\, -\mathbf{k}_4$ for the other graviton, we were able to target specific polarization information of the entangled state. In this way, we build a quantity $S$ computed by evaluating this 8-point correlation function at different observation angles $\theta, \phi, \theta'$ and $\phi'$ (keeping the choice of scalar momenta constant) that yields a value violating a Bell inequality.\\

The aim of this work was to, as proof of concept, design a Bell experiment out of the elements given by the simplest models of inflation. To this aim, we have computed only the polarization structure of a partially-disconnected contribution to the scalar 8-point function. It is left for future work to compute the integral over internal momenta and check if this is the dominant contribution to the full signal. Further, we acknowledge that an 8-point scalar correlation function is suppressed by a fourth power with respect to the two-point scalar correlation (the scalar power spectrum). However, this work points in the direction of searching for probes of the quantum nature of inflation in its most minimal form, and it is left for future works to find such probes with a more enhanced signal.

\section*{Acknowledgments}

The work of RJ is funded by the Simons Foundation. Funding for the work of PT, LS and RJ was partially provided by project PGC2018-098866- B-I00 y
FEDER “Una manera de hacer Europa”, and the “Center of Excellence Maria de Maeztu
2020-2023” award to the ICCUB (CEX2019- 000918-M) funded by MCIN/AEI/10.13039/501100011033
 DB acknowledges partial financial support from the COSMOS
network (www.cosmosnet.it) through the ASI
(Italian Space Agency) Grants 2016-24-H.0, 2016-24-H.1-2018 and
2020-9-HH.

\appendix

\section{Passive Vs active viewpoint of quantum state} \label{appendix: passive vs active}
In this appendix, we show that the "active viewpoint" and "passive viewpoints" for rotating the quantum state are equivalent in terms of the measured expectation values. The active viewpoint corresponds to rotating the quantum state $\ket{\Psi}$ by some angles $\theta,\phi$ while keeping the basis fixed, while the passive viewpoint corresponds to rotating the basis elements while keeping the state fixed. Technically the first diagonal entry of the density matrix $\rho_{\rm active}(\theta,\phi)$ in \eqref{eq: active density matrix} is the probability of getting $++$ for the actively rotated state, meanwhile the first entry of $\rho$ is the probability of getting the fixed state in the rotated basis. These probabilities are the same. The general proof follows:

\[
U = R(\theta)\otimes R(\phi)\,\,; 
\quad 
\rho_{\rm active} = U\,\rho_{\rm passive}\,U^\dagger \,\, ; 
\quad 
|e_i'\rangle = U\,|e_i\rangle,
\]
\begin{align}
(\rho_{\rm active})_{ii}
&= \langle e_i|\rho_{\rm active}|e_i\rangle
= \langle e_i|\,U\,\rho_{\rm passive}\,U^\dagger\,|e_i\rangle \notag \\[6pt]
&= \langle U^\dagger e_i|\rho_{\rm passive}|U^\dagger e_i\rangle
   \quad\bigl(\text{since }\langle a|U b\rangle=\langle U^\dagger a|b\rangle\bigr) \notag \\[6pt]
&= \langle e_i'|\rho_{\rm passive}|e_i'\rangle
   \quad\bigl(\text{because }U^\dagger|e_i'\rangle=|e_i\rangle\bigr)\ \notag \\[6pt]
&= (\rho_{\rm passive})_{\,ii}\,. \notag
\end{align}

In particular we have:
\begin{equation}
    \langle ++|\rho_{\rm active}|++\rangle
= \langle \theta,\phi|\rho_{\rm passive}|\theta,\phi\rangle\,, \notag
\end{equation}

which shows that every diagonal element of \(\rho_{\rm active}\) in the fixed \(\{+,\times\}\) basis equals the corresponding diagonal element of \(\rho\) in the rotated \(\{\theta,\phi\}\) basis.

\section{Computation of the 8-point function polarization structure}\label{appendix: 8-point function computation}

In this appendix, we show the calculation of the specific contribution to the expectation value of the 8-point scalar correlation function used in this work.

In Sec. \ref{subsec: our observable - 8 point func} we introduce the observable \eqref{eq: observable from 8-point} from the action \eqref{eq: graviton scalar scalar action}. We perform this calculation in the \textit{in-in} formalism (introduced in Sec. \ref{subsec: our observable - 8 point func}). Starting from expression \eqref{eq: 8-point exp val from braket} and expanding in powers of $H_I$:

\begin{align}
&\left._\text{in}\bra{\psi_a}\right. U(\eta,-\infty)\left(\prod_{n=1}^8\zeta_{\mathbf{k}_n} \right) U(-\infty,\eta)  \ket{\psi_a}_\text{in} \notag \\
 & \approx \bra{0}\, b_{\mathbf{p}_1}^{s_1}b_{\mathbf{p}_2}^{s_2}\, \tilde{\mathcal{T}}\Bigl\{
  \int_{-\infty}^\eta H_I\,\mathrm{d}\eta'\,
  \int_{-\infty}^\eta H_I\,\mathrm{d}\eta'
\Bigr\}
\, \left(\prod_{n=1}^8\zeta_{\mathbf{k}_n} \right) \notag \\
& \quad\,\,\,\mathcal{T}\Bigl\{
  \int_{-\infty}^\eta H_I\,\mathrm{d}\eta'\,
  \int_{-\infty}^\eta H_I\,\mathrm{d}\eta'
\Bigr\} \left(b_{\mathbf{p}_1}^{s_1}\right)^\dagger \left(b_{\mathbf{p}_2}^{s_2}\right)^\dagger   \ket{0}
\end{align}

where the interaction Hamiltonian is given by:

\begin{equation}
\label{interacting hamiltonian}
    H_I = -\frac{1}{2} \epsilon \, a^2\int\textup{d}^3\mathbf{x}\,\gamma^{ij}(\eta, \mathbf{x}) \partial_i \zeta(\eta,\mathbf{x}) \,\partial_j\zeta(\eta,\mathbf{x})
\end{equation}

We now use \eqref{interacting hamiltonian} to expand the time evolution operator. There are of course many terms, we focus for now on one where we expand both the forward and backward time evolution operators to second order. So we get:

\begin{widetext}
    \begin{align}
\label{8-point expansion}
\bra{0} b_{\mathbf{p}_1}^{s_1}b_{\mathbf{p}_2}^{s_2}\,& \tilde{\mathcal{T}}\Bigl\{
\int_{-\infty}^\eta \frac{1}{2} a^2\int\textup{d}^3\mathbf{x}\,\gamma^{ij}(\eta', \mathbf{x}) \partial_i \zeta(\eta',\mathbf{x}) \,\partial_j\zeta(\eta',\mathbf{x})\,\mathrm{d}\eta'\,\notag \\ 
  & \times \int_{-\infty}^\eta \frac{1}{2} a^2\int\textup{d}^3\mathbf{x}\,\gamma^{ij}(\eta', \mathbf{x}) \partial_i \zeta(\eta',\mathbf{x}) \,\partial_j\zeta(\eta',\mathbf{x})\,\mathrm{d}\eta'
\Bigr\}
 \,
\zeta_{\mathbf{k}_1}\zeta_{\mathbf{k}_2}\zeta_{\mathbf{k}_3}\zeta_{\mathbf{k}_4}
\zeta_{\mathbf{k}_5}\zeta_{\mathbf{k}_6}\zeta_{\mathbf{k}_7}\zeta_{\mathbf{k}_8}\, \notag \\
& \times 
\mathcal{T}\Bigl\{
  \int_{-\infty}^\eta \frac{1}{2} a^2\int\textup{d}^3\mathbf{x}\,\gamma^{ij}(\eta', \mathbf{x}) \partial_i \zeta(\eta',\mathbf{x}) \,\partial_j\zeta(\eta',\mathbf{x})\,\mathrm{d}\eta'\, \notag \\ 
  & \times \int_{-\infty}^\eta \frac{1}{2} a^2\int\textup{d}^3\mathbf{x}\,\gamma^{ij}(\eta', \mathbf{x}) \partial_i \zeta(\eta',\mathbf{x}) \,\partial_j\zeta(\eta',\mathbf{x})\,\mathrm{d}\eta'
\Bigr\} \bigl(b_{\mathbf{p}_1}^{s_1}\bigr)^\dagger\bigl(b_{\mathbf{p}_2}^{s_2}\bigr)^\dagger \ket{0}
\end{align}

\hfill \\

We will use the following expansions of the fields:

\begin{equation}
\zeta(\mathbf{x},\eta)
=\;
\int\!\frac{d^3k}{(2\pi)^3}\,
e^{\,i\mathbf{k}\cdot\mathbf{x}}
\bigl[a_{\mathbf{k}}\,U_k(\eta)
+a_{-\mathbf{k}}^\dagger\,U_k^*(\eta)\bigr]
\end{equation}

\begin{align}
\label{graviton field expansion}
\gamma_{ij}(\mathbf{x},\eta) = \sum_{s=+,\times}
\int\!\frac{d^3k}{(2\pi)^3}\,
e^{\,i\mathbf{k}\cdot\mathbf{x}}
\bigl[b_{\mathbf{k}}^s\,\epsilon_{ij}^s(\mathbf{k})\,\gamma_k(\eta)
+{\left(b_{-\mathbf{k}}^{s}\right)}^\dagger\,\epsilon_{ij}^{s*}(-\mathbf{k})\,\gamma_k^*(\eta)\bigr]
\end{align}
\hfill \\

and the following two point correlators:
\begin{equation*}
\bigl\langle \zeta_{\mathbf{k}}(\eta)\,\zeta_{\mathbf{p}}(\eta')\bigr\rangle
=
U_k(\eta)\,U_p^*(\eta')\,
\delta^{(3)}\!(\mathbf{k}+\mathbf{p}),
\end{equation*}

\begin{equation*}
\bigl\langle \gamma^s_{ij,\mathbf{k}}(\eta)\,\gamma^{s'}_{lm,\mathbf{p}}(\eta')\bigr\rangle
=
\gamma_k(\eta)\,\gamma_p^*(\eta')\,
\epsilon^s_{ij}(\mathbf{k})\,\epsilon^{s'}_{lm}(-\mathbf{p})\,
\delta_{ss'}\,
\delta^{(3)}\!(\mathbf{k}+\mathbf{p}),
\end{equation*}
where $U_k(\eta)$ and $\gamma_k(\eta)$ are the corresponding mode functions \cite{Mukhanov_2}, and $*$ denotes complex conjugation.
\end{widetext}

Now we contract \eqref{8-point expansion} by observing that each interaction Hamiltonian has two scalars and one graviton, so we can contract all of them with external operators/fields. We also set $\eta = 0$. This gives us the following:

\begin{widetext}
    \begin{align}
    &\propto\int_{-\infty}^0 d\eta_1 \int_{-\infty}^{\eta_1} d\eta_2  \int_{-\infty}^0 d\eta_3 \int_{-\infty}^{\eta_3} d\eta_4 \text{ }a^2(\eta_1)\,a^2(\eta_2)\,a^2(\eta_3)\,a^2(\eta_4)\text{ } \notag \\ 
    & \times \delta(\mathbf{k}_1+\mathbf{k}_2-\mathbf{p}_1)\delta(\mathbf{k}_5+\mathbf{k}_6+\mathbf{p}_1)\epsilon_{ij}^{s_1}(\mathbf{k}_{12}) k_1^i k_2^j \epsilon_{lm}^{s_1}(\mathbf{-k}_{56}) k_5^l k_6^m \gamma^*_{k_{12}}(\eta_1)
    \notag \\ 
    & \times \ U_{k_1}(\eta_1)\,U_{k_1}^*(0)\ \gamma^*_{k_{56}}(\eta_2)\ U_{k_2}(\eta_1)\,U_{k_2}^*(0)\ U_{k_5}(\eta_2)\,U_{k_5}^*(0)U_{k_6}(\eta_2)\,U_{k_6}^*(0)\ \notag \\
    &\times \delta(\mathbf{k}_3+\mathbf{k}_4-\mathbf{p}_2)\delta(\mathbf{k}_7+\mathbf{k}_8+\mathbf{p}_2)\epsilon_{lm}^{s_2}(\mathbf{-k}_{34}) k_3^l k_4^m\epsilon_{lm}^{s_2}(\mathbf{-k}_{78}) k_7^l k_8^m \gamma^*_{k_{34}}
    (\eta_3)\notag \\
    &\times \ U_{k_3}(\eta_3)\,U_{k_3}^*(0)\ \gamma^*_{k_{78}}(\eta_3)\ U_{k_4}(\eta_3)\,U_{k_4}^*(0)\ U_{k_7}(\eta_4)\,U_{k_7}^*(0)U_{k_8}(\eta_4)\,U_{k_8}^*(0)\,\,, \label{eq: appendix B relevant contribution of correlation}
\end{align}
\end{widetext}

up to constants. This corresponds to a partially disconnected diagram, and has a particular delta function structure which we take to be our case of interest. We now isolate the polarization terms:

\begin{widetext}
    \begin{multline}
\label{polarization structure}
    \propto \Bigg(\int_{-\infty}^0 d\eta_1 \int_{-\infty}^{\eta_1} d\eta_2  \int_{-\infty}^0 d\eta_3 \int_{-\infty}^{\eta_3} d\eta_4 \text{ }a^2(\eta_1)\, a^2(\eta_2)\, a^2(\eta_3)\, a^2(\eta_4)\text{ } \delta(\mathbf{k}_1+\mathbf{k}_2-\mathbf{p}_1)\delta(\mathbf{k}_5+\mathbf{k}_6-\mathbf{p}_2) \\
    \times \gamma^*_{k_{12}}(\eta_1)
    \ U_{k_1}(\eta_1)\,U_{k_1}^*(0)\ \gamma^*_{k_{56}}(\eta_2)\ U_{k_2}(\eta_1)\,U_{k_2}^*(0)\ U_{k_5}(\eta_2)\,U_{k_5}^*(0)U_{k_6}(\eta_2)\,U_{k_6}^*(0)\ 
    \delta(\mathbf{k}_3+\mathbf{k}_4+\mathbf{p}_1)\\\times\delta(\mathbf{k}_7+\mathbf{k}_8+\mathbf{p}_2) \gamma^*_{k_{34}}
    (\eta_3)\ U_{k_3}(\eta_3)\,U_{k_3}^*(0)\ \gamma^*_{k_{78}}(\eta_3)\ U_{k_4}(\eta_3)\,U_{k_4}^*(0)\ U_{k_7}(\eta_4)\,U_{k_7}^*(0)U_{k_8}(\eta_4)\,U_{k_8}^*(0)\Bigg)\\
    \times \epsilon_{ij}^{s_1}(\mathbf{k}_{12}) k_1^i k_2^j\epsilon_{ij}^{s_1}(\mathbf{-k}_{56}) k_5^i k_6^j\epsilon_{lm}^{s_2}(\mathbf{k}_{34}) k_3^l k_4^m\epsilon_{lm}^{s_2}(\mathbf{-k}_{78}) k_7^l k_8^m\\
    \equiv  F(\mathbf{k}_1,\mathbf{k}_2,\mathbf{k}_3,\mathbf{k}_4,\mathbf{k}_5,\mathbf{k}_6,\mathbf{k}_7,\mathbf{k}_8)  \times \epsilon_{ij}^{s_1}(\mathbf{k}_{12}) k_1^i k_2^j\epsilon_{ij}^{s_1}(\mathbf{-k}_{56}) k_5^i k_6^j\epsilon_{lm}^{s_2}(\mathbf{k}_{34}) k_3^l k_4^m\epsilon_{lm}^{s_2}(\mathbf{-k}_{78}) k_7^l k_8^m\ \\
    \vspace{0.7cm}
\end{multline}
\end{widetext}

So we have isolated the polarization terms to be outside of the time integrals, now if we take $\mathbf{k}_5 = -\mathbf{k}_1$, $\mathbf{k}_6 = -\mathbf{k}_2$, $\mathbf{k}_7 = -\mathbf{k}_3$, $\mathbf{k}_8 = -\mathbf{k}_4$ we get:

\begin{align}
   & F(\mathbf{k}_1,\mathbf{k}_2,\mathbf{k}_3,\mathbf{k}_4,\mathbf{-k}_1,\mathbf{-k}_2,\mathbf{-k}_3,\mathbf{-k}_4) \notag \\
   & \hspace{0.9cm} \times \big(\epsilon_{ij}^{s_1}(\mathbf{k}_{12}) k_1^i k_2^j \epsilon_{lm}^{s_2}(\mathbf{k}_{34}) k_3^l k_4^m \big)^2
\end{align}

This is the relevant polarization structure that we use in Sec. \ref{subsec: CHSH ineq from 8-point} to design a Bell-violating inequality from the 8-point function as our observable today. This expression is valid for any particular choice of the polarizations $s_1$ and $s_2$. Note here $s_1$ and $s_2$ refer to the polarization states of the two gravitons.  \\

One may not find the above derivation satisfactory because it is only true for one specific contraction/expansion. For instance one will also have terms in which one can expand $U(\eta)$ to zeroth order and $U^{\dagger}(\eta)$ to fourth order or $U(\eta)$ to first order and $U^{\dagger}(\eta)$ to third order, etc. In fact we can show that the form \eqref{polarization structure} is generic for the given delta function structure. This comes from the following observation: any term coming from the generic correlator in \eqref{8-point expansion} that has the specific delta function structure of
\begin{equation}
\label{eqap: relevant delta structure of calc}
    \delta(\mathbf{k}_1 +~ \mathbf{k}_2-~\mathbf{p}_1)\delta(\mathbf{k}_5 +~ \mathbf{k}_6+\mathbf{p}_1)\delta(\mathbf{k}_3 +~ \mathbf{k}_4 -~ \mathbf{p}_2)\delta(\mathbf{k}_7+\mathbf{k}_8+\mathbf{p}_2)
\end{equation}

necessarily has the polarization structure of

\begin{equation}
\label{eqap: relevant polariz structure}
    \epsilon_{ij}^{s_1}(\mathbf{k}_{12}) k_1^i k_2^j\epsilon_{ij}^{s_1}(\mathbf{-k}_{56}) k_5^i k_6^j\epsilon_{lm}^{s_2}(\mathbf{k}_{34}) k_3^l k_4^m\epsilon_{lm}^{s_2}(\mathbf{-k}_{78}) k_7^l k_8^m\,\, .
\end{equation}

To show this, let us focus on a single delta function, for instance $\delta(\mathbf{k}_1+\mathbf{k}_2-\mathbf{p}_1)$. Then this delta function comes from contracting $H_I\propto \epsilon_{ij}^{s_1}(\mathbf{q})\partial_i\zeta_2  \partial_j\zeta_1$ with either $\zeta_{k_1}\zeta_{k_2}(b_{\mathbf{p}_1}^{s_1})^{\dagger}$ or $\zeta_{k_1}\zeta_{k_2} b_{\mathbf{p}_1}^{s_1}$. So we have 4 possibilities in total, since $\zeta_1$ can contract with $\zeta_{k_1}$ or $\zeta_{k_2}$ (which forces $\zeta_2$ to contract with the other one). Now, since $\epsilon_{ij}$ is symmetric, the above two possibilities give the same answer. Finally, to distinguish whether $\epsilon_{ij}$ contracts with $b_{\mathbf{p}_1}^{s_1}$ or with $(b_{\mathbf{p}_1}^{s_1})^{\dagger}$ we note that the minus sign in the delta function characterizes which one and it is easily seen from \eqref{graviton field expansion} that the delta function implies the contraction is made with $(b_{\mathbf{p}_1}^{s_1})^{\dagger}$ rather than with $b_{\mathbf{p}_1}^{s_1}$, implying that the delta function $\delta(\mathbf{k}_1+\mathbf{k}_2-\mathbf{p}_1)$ necessarily implies the existence of the specific polarization term $\epsilon_{ij}^{s_2}(\mathbf{k}_{12}) k_1^i k_2^j$. By the exact same logic, one gets all the other polarization terms.

\bibliographystyle{unsrt}

\end{document}